\documentclass[aps,prb,twocolumn,superscriptaddress,showpacs,amsmath,amssymb,floatfix]{revtex4}

\usepackage{graphicx}          
\usepackage{dcolumn}           
\usepackage{colordvi}          
\usepackage{ifthen}            
\usepackage{bm}                

\usepackage{german}            
\newcommand{\art}[6]{\bibitem{#1} #2, #3 {\bf #4}, #5 {(#6)}.}%
\newcommand{\book}[4]{\bibitem{#1} #2, {\it #3} {(#4)}.}%
\newcommand{\info}[2]{\bibitem{#1} #2}%
\newcommand{\aplx}{Appl. Phys. Lett.}

\newcommand{\prax}{Phys. Rev. A}

\newcommand{\prlx}{Phys. Rev. Lett.}
\newcommand{\oex}{Opt. Express}
\newcommand{\scix}{Science}
\newcommand{\natx}{Nature}
\newcommand{\natpx}{Nature Photon.}

\newcommand{\jqex}{IEEE J. Quantum Electron.}

\newcommand{\mm}{$\mu$m}

\newcommand{\Sem}{Scanning electron micrograph}

\newcommand{\fdtd}{finite-difference time-domain}

\newcommand{\Tamosiunas}{$\rm Tamo\check{s}i\bar{u}nas$}

\begin{document}
\clearpage
\title{Electrically Switchable Photonic Molecule Laser}
\author{Gernot Fasching}%
\affiliation{%
Photonics Institute, Vienna University of Technology, Gusshausstrasse 27/387, 1040 Wien, Austria%
}%
\affiliation{%
Centre for Micro- and Nanostructures, Vienna University of Technology, Florag. 7, 1040 Wien, Austria%
}%
\author{Christoph Deutsch}
\author{Alexander Benz}
\affiliation{%
Photonics Institute, Vienna University of Technology, Gusshausstrasse 27/387, 1040 Wien, Austria%
}%
\affiliation{%
Centre for Micro- and Nanostructures, Vienna University of Technology, Florag. 7, 1040 Wien, Austria%
}%
\author{Aaron Maxwell Andrews}
\author{Pavel Klang}
\affiliation{%
Institute of Solid-State Electronics, Vienna University of Technology, Florag. 7, 1040 Wien, Austria%
}%
\affiliation{%
Centre for Micro- and Nanostructures, Vienna University of Technology, Florag. 7, 1040 Wien, Austria%
}%
\author{Reinhard Zobl}
\affiliation{%
Institute of Solid-State Electronics, Vienna University of Technology, Florag. 7, 1040 Wien, Austria%
}%
\author{Werner Schrenk}
\author{Gottfried Strasser}
\affiliation{%
Institute of Solid-State Electronics, Vienna University of Technology, Florag. 7, 1040 Wien, Austria%
}%
\affiliation{%
Centre for Micro- and Nanostructures, Vienna University of Technology, Florag. 7, 1040 Wien, Austria%
}%
\author{Paulius Ragulis}
\author{Vincas \Tamosiunas}
\affiliation{%
Semiconductor Physics Institute, A. Gostauto 11, 01108 Vilnius,
Lithuania}%
\author{Karl Unterrainer}{
\affiliation{%
Photonics Institute, Vienna University of Technology, Gusshausstrasse 27/387, 1040 Wien, Austria%
}%
\affiliation{%
Centre for Micro- and Nanostructures, Vienna University of Technology, Florag. 7, 1040 Wien, Austria%
}%
%
\begin{abstract}
We have studied the coherent intercavity coupling of the
evanescent fields of the whispering gallery modes of two terahertz
quantum-cascade lasers implemented as microdisk cavities. The
electrically pumped single-mode operating microcavities allow to
electrically control the coherent mode coupling for proximity
distances of the cavities up to 30-40 \mm. The optical emission of
the strongest coupled photonic molecule can be perfectly switched
by the electrical modulation of only one of the coupled
microdisks. The threshold characteristics of the strongest coupled
photonic molecule demonstrates the linear dependence of the gain
of a quantum-cascade laser on the applied electric field.
\end{abstract}
\maketitle

In general, photons couple much less with respect to electrons
allowing photonic circuits to be faster, providing more bandwidth
and having lower power consumption compared to their electronic
counterparts. Merging optics and electronics for future optical
circuits \cite{Bin07, Ozb06} is still an open challenge mainly due
to the lack of a chip based platform providing photon sources and
manipulation units at once. Hybrid solutions are on the way
\cite{Yar07, Par08} to bridge between sources and circuits, but
the efficient coupling into photonic circuits remains still a
problem to be overcome. The III-V material system is well suited
for fully functional optoelectronics, as it combines both optical
and electronic functions. GaAs based compact and unipolar
mid-infrared \cite{Fai94} and terahertz (THz) \cite{Koe02}
emitting lasers called quantum-cascade lasers (QCLs) have already
been realized. The light amplification is based on intersubband
(ISB) transitions, i.e. transitions between one-dimensional
quantized energy levels within one band of a cascaded
semiconductor heterostructure. Quantum-cascade microlasers with
in-plane highly directional \cite{Gma98}, in-plane unidirectional
\cite{Fas05}, or surface \cite{Mah08, Muj09} emission as well as
active photonic crystals \cite{Zha07, Cha09, Ben09} have also been
established, providing the basic building blocks for light
manipulation and waveguiding. Hence, photon sources and waveguides
are already established in the THz spectral range, but the optical
intercavity coupling has not been studied so far. \newline\noindent%
In this Letter we investigate for the first time the electrically
controlled coherent optical coupling between two
whispering-gallery modes (WGMs), which is based on phase and mode
matched power exchange of the constituting fields. Hence, by the
linear combination of the optical fields of microdisk lasers one
can create so called photonic molecules \cite{Bay98} (PMs).
%
Compared to former realizations of PMs, e.g. microspheres
\cite{Muk99,Rak04} or semiconductor microcavities
\cite{Bay98,Nak05}, PMs based on microdisk THz-QCLs differ in
several important aspects from these approaches. First, THz-PMs
comprise an electrically pumped optical gain enabling fast
electrically controlled mode tuning and switching. Second, the
plasmonic mode confinement permits sub-wavelength sized cavities.
This allows to study the optical coupling of just two single modes
and the precise control over the resonance frequency. And third,
the combination of gain switching and exact spatial cavity
configuration offers a precise control over the resulting mode
configuration and on chip integration.\newline\noindent%
The investigated PMs consist of two THz-QCLs having each a
cylindrical resonator as shown in Fig. \ref{f:SEM}, i.e. a
microdisk cavity, with the same nominal dimensions of radius
$R$=45 \mm\ and height $H$=16.2 \mm. The varying proximate
spacings between the microdisks result in different coupling
strengths. The proximate spacing $x$ will be included in the
terminology as PM-$x$. \newline\noindent %
\begin{figure}[tb!]
\resizebox{8cm}{!}{\includegraphics{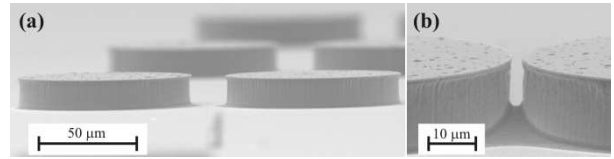}}
\caption{%
\Sem\ images of double-metal microdisks forming electrically
tunable THz photonic molecules with a proximity spacing of (a) 12
(PM-12) and (b) 2 \mm\ (PM-2).%
}\label{f:SEM}%
\end{figure}
The active gain region of the microdisks is based on a four
well/barrier GaAs$/Al_{0.15}Ga_{0.85}As$ heterostructure
\cite{Wil03}, which has been repeated 271 times during the growth
to achieve a large modal gain. The heterostructure is embedded
between two metal layers serving as electrical contacts and
allowing for strong vertical mode confinement ($\lambda \gg H$),
whereas the lateral confinement is provided by the impedance
mismatch between the gain material and the air. Details of the
design, growth, and processing can be found elsewhere
\cite{Fas05}. The field dependent emitted optical power of the PMs
has been measured in a closed light pipe equipped with a Ga doped
Ge (Ge:Ga) detector. The PMs were operated in pulsed-mode with 100
ns short pulses and double modulated at 100 Hz with a duty-cycle
of 50 \% to allow for the detection with a Ge:Ga detector. The
light pipe was immersed in liquid helium, such that room
temperature background radiation
could not distort the measurements. \newline\noindent%
\begin{figure}[tb!]
\resizebox{7cm}{!}{\includegraphics{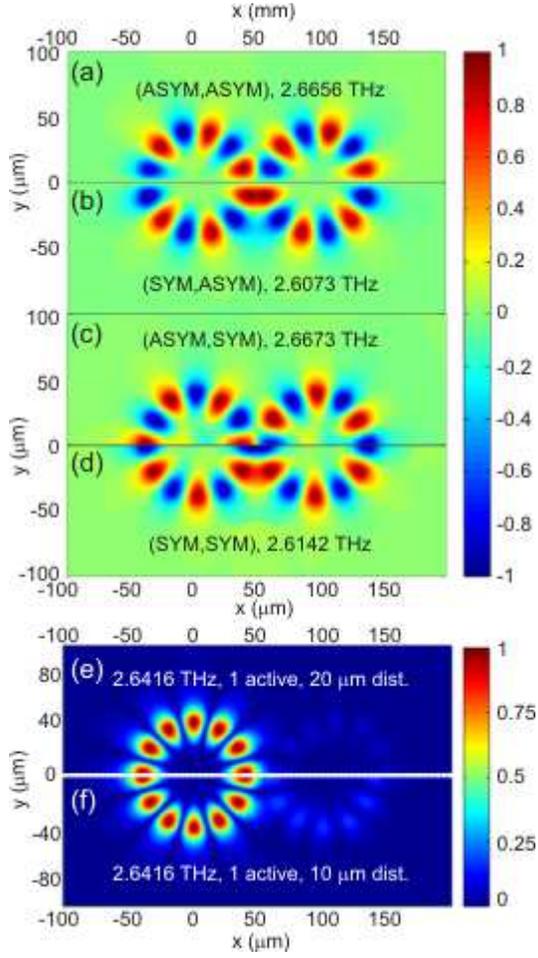}}
\caption{%
Spatial field distributions of the $z$ component of the electric
field. (a-d) The sign of the real part of the amplitude is
assigned to the absolute value of the amplitude in order to
indicate the phase differences of the restricted symmetric and
antisymmetric simulations in the case of a PM-2. (e,f) Full area
simulations of a PM-20 and a PM-10 having only one active
microdisk. In both molecules the lasing mode couples into the
passive cavity.%
}\label{f:FDTDmodeprofiles}%
\end{figure}
Full 3-D \fdtd\ (FDTD) simulations have been performed to reveal
the properties of PM modes. A custom-made code \cite{Tam05} was
employed for this purpose, which includes several enhancements
over the classical FDTD, such as a Maxwell-Bloch module for
two-level quantum systems. In addition to full-window simulations
\cite{Inf1}, higher resolution simulations were performed with
restricted symmetry to reveal the detailed field configurations.
The modes are labelled by a pair ($x,y$), indicating a symmetric
(SYM) or antisymmetric (ASYM) field configuration with respect to
the symmetry planes perpendicular ($x$) or parallel ($y$) to the
molecule axis. ($ASYM,y$) modes are {\it pushed} further inside
the resonator compared to ($SYM,y$) modes, resulting in different
effective optical paths causing higher emission frequencies for
the ($ASYM,y$) modes as shown in Fig.
\ref{f:FDTDmodeprofiles}(a-d). For the case of one active and one
passive/absorbing microdisk as shown for a PM-20 and a PM-10 in
Fig. \ref{f:FDTDmodeprofiles}(e,f), the mode is well localized in
the active cavity but couples also into the passive cavity. Hence,
the operation of a PM is tunable by
changing the effective loss of the passive cavity.\newline\noindent%
\begin{figure}[tb!]
\resizebox{7cm}{!}{\includegraphics{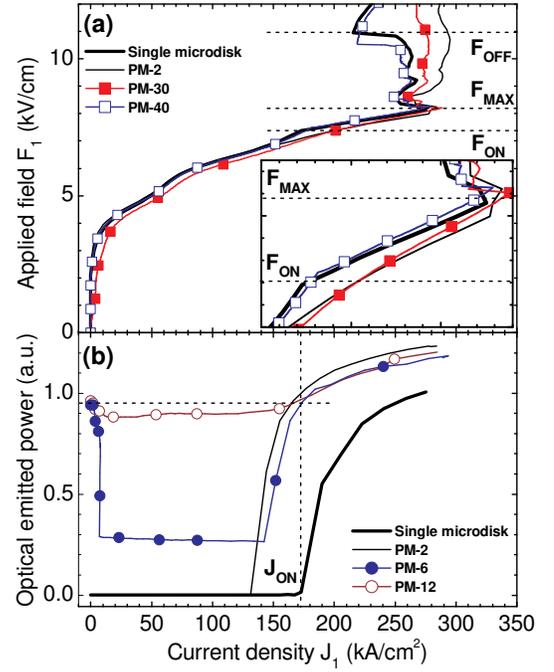}}
\caption{
(a) The current density-applied field ($J-F$) characteristics of a
single microdisk and several PMs at T=5 K. The applied electric
fields at the lasing onset, at the maximum lasing emission, and at
the end of the lasing emission of a single microdisk are labelled
$F_{ON}$, $F_{MAX}$ and $F_{OFF}$, respectively. The applied field
of the second microdisk of each PM was kept constant at
$F_{const}$=($F_{ON}+F_{MAX})/2$. The inset depicts a blow up of
the $J-F$ characteristics between $F_{ON}$ and $F_{MAX}$ to
clarify the increase in current for the PM-2 and the PM-30 within
this region. (b) The current density-optical emitted power
characteristics of a single microdisk and several PMs at T=5 K.
The dashed horizontal line mark the lasing emission of the PM-6
and PM-12 for $J_1$=0. The dashed vertical line mark the threshold
current density $J_{ON}$ of a single microdisk.
}\label{f:LIV}%
\end{figure}
The optical intercavity coupling between microdisks forming a PM
has a strong impact on the electrical and optical characteristics
as shown in Fig. \ref{f:LIV}. A single, uncoupled microdisk
exhibits the typical current density-applied field characteristics
due to ISB tunneling and lasing emission \cite{Sir98}, i.e. a kink
at the onset of the lasing emission and a negative differential
resistivity regime above the maximum lasing emission labelled
$F_{ON}$ and $F_{MAX}$ in Fig. \ref{f:LIV}(a), respectively. %
\begin{figure*}[t!]
\resizebox{12cm}{!}{\includegraphics{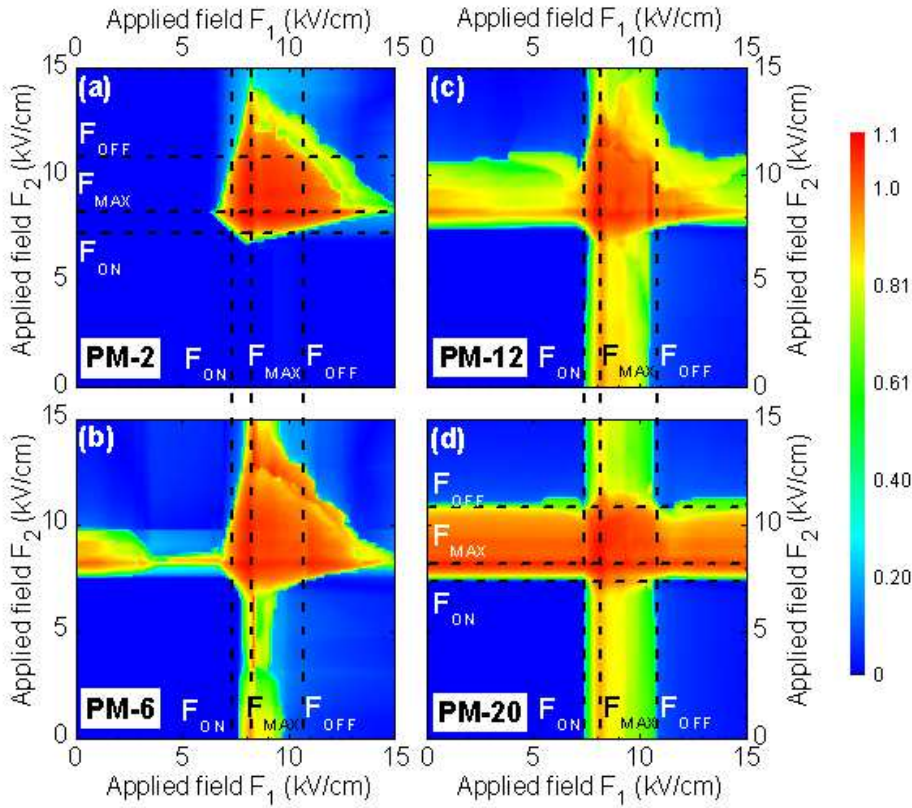}}
\caption{%
The optical emitted power of four PMs as a function of the applied
electric fields with decreasing intercavity coupling from (a) to
(d). The dashed horizontal and vertical lines labelled $F_{ON}$,
$F_{MAX}$, and  $F_{OFF}$, respectively, mark the applied electric
fields at the lasing onset, the maximum lasing emission, and at
the end of the lasing emission of a single, uncoupled microdisk.
}\label{f:ContourIntensity}%
\end{figure*}
Compared to that, the PM-2 and the PM-30 exhibit higher current
densities starting below $F_{ON}$ up to well above $F_{OFF}$. This
is a consequence of the strong {\it optical} intercavity coupling
causing the additional photon enhanced current flow. The PM-40
does not show such an increase indicating that the
strong optical coupling is limited to $\approx$30-40 \mm.\newline\noindent%
Fig. \ref{f:LIV}(b) shows the optical emitted power of a single
microdisk and several PMs. The PM-2 depicts the same qualitative
behaviour and the same threshold current density $J_{ON}$ of a
single microdisk, i.e. $J_1+J_2=2 J_{ON}$, reflecting that a PM-2
acts
as a {\it single} cavity QCL.\newline\noindent%
The PM-6 and the PM-12 exhibit emission already at $J_1=0$, i.e.
the single peak gain of the second microdisk is larger than the
loss of the coupled system. The emission of these PMs is
decreasing around $J_1\approx 10$ A/cm$^{2}$ due to carrier
injection into the other microdisk leading to ISB absorption. The
lasing emission of these PMs recover exactly at $J_1$ = $J_{ON}$,
i.e. the additional loss is balanced by the gain.\newline\noindent%
Fig. \ref{f:ContourIntensity} shows the electrical control of the
lasing emission of four PMs. The unique possibility to externally
control the gain/loss of each cavity of a PM allows to tune the
spatial mode configuration. The strongest coupled system shown in
Fig. \ref{f:ContourIntensity}(a) lases only if both cavities
provide gain. This electrically controlled purely optical
conditional switching represents a logical AND operation. All
weaker coupled systems exhibit lasing also if only one microdisk
is biased beyond the lasing threshold field $F_{ON}$, which is an
equivalent to a logical OR operation. Adding a NOT operation by
electrically inverting the applied field of a single microdisk
completes the set of necessary basic operations, i.e. a logical
AND, OR, and NOT, to perform all possible Boolean operations. This
demonstrates, that a network of optically coupled THz laser can be
utilized to perform any complex logical operation.\newline\noindent%
The general transition from the strongly to the weakly coupled PM
by increasing the spacing between the microdisks from 2 to 20 \mm\
is directly reflected by the change from the
trapezoid-shaped to the cross-shaped lasing emission.\newline\noindent%
The onset of the optical emission of the PM-2 is linearly and
equally dependent on the applied electric fields $F_1$ and $F_2$
below $F_{MAX}$, i.e. below the onset of the negative differential
resistivity regime. Hence, (i) the gain of a {\it single} THz-QCL
is linear proportional to the applied field above as well as below
the threshold with the {\it same} slope \cite{Kro07} and (ii) the
alignment of the energy levels is achieved well before the lasing
threshold field $F_{ON}$. The onset of the gain reduction beyond
$F_{MAX}$ requires both applied fields to increase linearly to
maintain lasing emission. Thus, the gain of a {\it single} THz-QCL
is also linear proportional to the applied field above $F_{MAX}$
{\bf , i.e.} in the negative differential resistivity regime. The
lasing emission of the PM-2 molecule extends beyond $F_{OFF}$ and
is limited by the onset of the gain reduction in the other
microdisk
at $F_{MAX}$.\newline\noindent%
The decrease in coupling efficiency allows for additional lasing
regions between $F_{ON}$ and $F_{OFF}$ as shown in Fig.
\ref{f:ContourIntensity}(b-d). The lasing emission in Fig.
\ref{f:ContourIntensity}(b) is strongly dependent for $F_1\approx$
3.7 kV/cm. As discussed in Fig. \ref{f:LIV}(b), the injected
carriers in the other microdisk cause an increase in absorption of
the optically coupled system. The increase of the spacing to 12
\mm\ leads to a pronounced increase of the single microdisk lasing
areas extending the available lasing range. Finally, the optical
emission of the PM-20 shown in Fig. \ref{f:ContourIntensity}(d)
covers almost only the whole applied electric field range of two
independent microdisks between $F_{ON}$ and $F_{OFF}$. Hence, the
crossed-like emission indicates the strong reduction of the mutual
coupling.\newline\noindent%
In summary, we have demonstrated the coherent evanescent coupling
of two THz emitting microdisk lasers representing a PM. %
The electrical characteristics demonstrate mutual coupling up to
30-40 \mm\ intercavity spacing which is comparable to the
intracavity lasing wavelength. As a consequence of carrier
lifetimes in the picosecond range THz-PMs can be operated with 100
ns short pulses performing extremely fast electrically gated
optical modulation as well as logical AND and OR operations. The
characteristics of the optical emitted power demonstrate a linear
control of the ISB gain by the applied electric field. \newline\noindent%
THz emitting PMs might have a great potential for sub-wavelength
photonic devices which can be easily realized by standard
lithography. Hence, they might find their way into photonics as
applications for highly sensitive gas sensors using the evanescent
coupled fields with electrical readout or coherently coupled laser
arrays with vertical outcoupling for power emission and possible
beam shaping and steering. On chip integration of these
electrically controllable discrete elements as couplers, filters,
gates or flip-flops might allow the realization of complex
plasmonic
circuits.\newline\noindent%
The authors acknowledge financial support by the Austrian Science
Fund (SFB-ADLIS, SFB-IRON), the Austrian Nano Initiative project
(PLATON), the EC Program ``POISE'' (TRM), the Lithuanian State and
Studies Foundation (contract No C-07004), and the Society for
Microelectronics (GME, Austria).%
   \bibliographystyle{plain}

\end{document}